\documentclass[12pt]{iopart}
\usepackage{amssymb,amsfonts}
\newcommand{\Zint}{\mathbb{Z}}
\newcommand{\Real}{\mathbb{R}}
\newcommand{\irrep}[1]{\bi{#1}}
\newcommand{\eis}[3]{~\ensuremath{{\cal E}^{#1}_{\irrep{#2};#3}}}
\newcommand{\exc}{\ensuremath{E_{d+1(d+1)}}}

\newcommand{\R}{\mathcal{R}}
\newcommand{\F}{\mathcal{F}}

\newcommand{\M}{\mathcal{M}}

\newcommand{\T}{\mathcal{T}}
\newcommand{\Rank}{{\rm Rank}\ }
\newcommand{\vect}{\ensuremath{{\bf V}}}
\newcommand{\spi}{\ensuremath{{\bf S}}}
\newcommand{\spb}{\ensuremath{{\bf C}}}
\def\pa{\partial}
\def\sp{\quad , \quad}
\begin{document}
% Journal identifier can be put here if required, e.g.
%\jl{14}
\begin{flushright} NORDITA-1999/65 HE \\
NBI-HE-99-40\\ HUTP-99/054 \\ LPTHE-99-34 
 \\ {\tt hep-th/9910115} \end{flushright}
  \vfill
\title[Eisenstein Series in String Theory]
{Eisenstein Series in String Theory${}^{\star}$}

\author{Niels A. Obers \dag %\footnote{obers@nordita.dk}
 \ and Boris Pioline \ddag %\footnote{pioline@ulam.harvard.edu}
$^{\ast}$}

\address{\dag\ Nordita and NBI, Blegdamsvej 17, DK-2100 Copenhagen,
Denmark\\{\tt \ \  obers@nordita.dk} }

\address{\ddag\
Jefferson Physical Laboratory, Harvard University, Cambridge MA
02138, USA\\ {\tt \ \ pioline@ulam.harvard.edu} }

\begin{abstract}
We discuss the relevance of Eisenstein series for representing
certain $G(\Zint)$-invariant string theory amplitudes which
receive corrections from BPS states only. The Eisenstein series
are constructed using $G(\Zint)$-invariant mass formulae and are
manifestly invariant modular functions on the symmetric space
$K\backslash G(\Real)$ of non-compact type, with $K$ the maximal
compact subgroup of $G(\Real)$. In particular, we show how
Eisenstein series of the T-duality group $SO(d,d,\Zint)$ can be
used to represent one- and $g$-loop amplitudes in compactified
string theory. We also obtain their non-perturbative extensions in
terms of the Eisenstein series of the U-duality group $\exc
(\Zint)$.
\end{abstract}
 \vfill

  \begin{flushleft} \small
  ${}^{\star}$ Talk presented by N.O. at Strings '99, Potsdam,
Germany (July 19-24, 1999). Work supported in part by TMR networks
ERBFMRXCT96-0045 and ERBFMRXCT96-0090. \\ $^{\ast}$ \small On
leave of absence from LPTHE, Universit{\'e} Pierre et Marie Curie,
PARIS VI and Universit{\'e} Denis Diderot, PARIS VII, Bo\^{\i}te
126, Tour 16, 1$^{\it er}$ {\'e}tage, 4 place Jussieu, F-75252
Paris CEDEX 05, FRANCE
  \end{flushleft}
%\submitted

\maketitle
\section{Introduction}

Worldsheet modular invariance has played a major role in the
construction of consistent anomaly-free perturbative string
models. At the one-loop level, it demands that the string partition
function be invariant under the $Sl(2,\Zint)$ fractional linear
transformations of the modular parameter of
world-sheet torus. The advent of target-space and non-perturbative
dualities has brought into play yet another branch of the
mathematics of automorphic forms invariant under infinite discrete
groups $G(\Zint)$. These dualities arise in models with
many supersymmetries, where the scalar fields take values in
a rigid symmetric space $K\backslash G(\Real)$,
where  $K$ is the maximal compact subgroup of $G$. Duality
symmetries identify points in $K\backslash G(\Real)$ differing by the right action
of an infinite discrete subgroup $G(\Zint)$ of $G(\Real)$, and all
physical amplitudes
are constrained to be invariant under the transformations in $G(\Zint)$.
This includes in particular the mapping class group $Sl(d,\Zint)$
in the case of toroidal compactifications of
diffeomorphism-invariant theories, the T-duality group
$SO(d,d,\Zint)$ in toroidal compactifications of string theories,
as well as the non-perturbative U-duality group $\exc(\Zint)$ in
maximally supersymmetric compactified M-theory  or non-perturbative
type II string theory
\cite{Hull:1995mz,Townsend:1995kk,Witten:1995ex} (see for instance
\cite{Giveon:1994fu,Obers:1998fb} for reviews and exhaustive list
of references).  These duality symmetries are to be contrasted
with dualities between distinct string
theories, which, although less tied to supersymmetry, yield
much weaker constraints.

The invariance under duality symmetries, although valid for all
physical amplitudes, is usually not sufficient to completely
determine them. For a certain class of ``BPS saturated''
amplitudes however, supersymmetry gives further constraints, in
the form of second order differential equations in the most
favorable half-BPS saturated case
\cite{Pioline:1998mn,Paban:1998qy,Green:1998by}. Combining these
constraints with duality invariance and behavior at the boundary
of the moduli space allows in many cases to determine exact
non-perturbative results not obtainable otherwise, which can then
be analyzed at weak coupling to provide interesting insights on
instanton calculus in string theory\footnote{Exact amplitudes can
also be obtained from string dualities, and yield valuable
insights too, see for instance
\cite{Harvey:1996ir,Bachas:1997mc,Gregori:1997hi,Antoniadis:1997zt}.}
\cite{Green:1997tv,Kiritsis:1997em,Obers:1999um} (see also Ref.
\cite{Kiritsis:1999ss} for lecture notes). As an illustration,
\Tref{tab1}  lists for a number of compactifications the
(non-)perturbative symmetries that apply to all amplitudes along
with the amount of supersymmetry of the theory and a selection of
1/2 and 1/4-BPS amplitudes.

The prototypical example of such an exactly computable coupling is
the $R^4$ coupling in ten dimensional type IIB theory. This was
conjectured \cite{Green:1997tv} and later proved
\cite{Berkovits:1997pj,Pioline:1998mn,Green:1998by} to be a
certain (non-holomorphic) automorphic form of weight 0 of the S-duality group
$Sl(2,\Zint)$, namely the Eisenstein series of order 3/2
\begin{eqnarray}
\label{r4iib} f_{R^4}^{IIB}&=&\frac{1}{ l_P^{2}}
\eis{Sl(2,\Zint)}{2}{s=3/2}\ ,\\ \eis{Sl(2,\Zint)}{2}{s}\ &=&
\sum_{(m,n)\ne 0}  \left[ \frac{\tau_2}{|m+n\tau|^2} \right]^{s}
=\zeta(2s) \sum_{(p,q)=1} \frac{
 \tau_2^s}{|p+q\tau|^2} \ ,  \label{sl2eis}
\end{eqnarray}
where $\tau=a+i/g_s$ is the complexified string coupling, parametrizing the
upper-half-plane $U(1)\backslash Sl(2,\Real)$, and $l_P$ the
ten-dimensional S-duality invariant Planck length $l_P =g_{s}^{1/4} l_s$.
By expanding this series at weak coupling $\tau_2\to\infty$, one recovers the
known tree-level and one-loop terms, along
with an infinite series of exponentially suppressed terms of order $e^{-N/g_s}$,
attributed to $N$ D-instanton configurations. It is easy to check that the
above Eisenstein series is an eigenmode of the Laplacian on the upper
half plane,
\begin{equation}
\label{lsl2} \fl
\Delta_{U(1)\backslash Sl(2)}
\eis{Sl(2,\Zint)}{2}{s} = \frac{s(s-1)}{2} \eis{Sl(2,\Zint)}{2}{s}
\ ,\qquad \Delta_{U(1)\backslash Sl(2)}=
\frac{1}{2}\tau_2^2(\partial_{\tau_1}^2+\partial_{\tau_2}^2)\ .
\end{equation}
Supersymmetry on the other hand requires the
exact $R^4$ coupling to be an eigenmode of the same Laplacian
with the precise eigenvalue $3/8$, which together with the assumption of
at most polynomial growth at weak coupling uniquely selects the
Eisenstein series of order $3/2$.

In this talk we discuss the construction and utility of
automorphic forms of the duality groups $G(\Zint)$ generalizing
the Eisenstein series of $Sl(2,\Zint)$ (see Ref.
\cite{Obers:1999um} for an extensive treatment). For the
representation theory of these groups, in particular the
multiplets of half-BPS states that we will use, we refer the
reader to the review  \cite{Obers:1998fb} and its very brief
summary \cite{Obers:1998rn}. Section 2 presents the general
construction of Eisenstein series, Section 3 its applications to
perturbative string amplitudes; Section 4 extends these results to
the full non-perturbative expression and finally Section 5
mentions some open directions.

\begin{table}
\begin{center}
\begin{tabular}{|c|c|c|c|c|} \hline \hline theory  & symmetry &SUSY & 1/2 BPS & 1/4 BPS \\ \hline
IIB    & $  Sl(2,\Zint) $ & 32 & $R^4$, $\lambda^{16}$  & $R^6, \lambda^{24}$
\\ \hline { M/}$ T^{d+1}$  ($\sim$  II/$ T^{d}$) &  $
\exc (\Zint)$ & 32 & $R^4$, $\lambda^{16}$  & $R^6, \lambda^{24}$
\\  \hline  Het$ /T^n \;\,(n \leq 5) $  &  $ SO(n,16+n,\Zint) $ & 16 & $R^2,F^4, \lambda^{8}$ & $F^6, \lambda^{12}$
\\  \hline  Het$ /T^6 $  &  $ SO(6,22,\Zint) \times Sl(2,\Zint)$ & 16 & $R^2,F^4, \lambda^{8}$ & $F^6, \lambda^{12}$
\\  \hline  Het$ /T^7 $  &  $ SO(8,24,\Zint)$ & 16 & $R^2,F^4, \lambda^{8}$ & $F^6, \lambda^{12}$
\\ \hline   IIB/$ K_3$  & $ SO(5,21,\Zint)$  & 16 & $R^2,H^4,\lambda^8$ &$H^6,\lambda^{12}$\\ \hline
\end{tabular}
\caption{(Non-)perturbative symmetries of various compactified string
theories. \label{tab1}}
\end{center}
\end{table}

\section{Generalized Eisenstein series}
We focus on string or M-theory models whose moduli space is spanned by
a symmetric space of non-compact type
\begin{equation}
\label{iwasawa}
  K \backslash G(\Real) \sp G(\Real) = K \cdot A \cdot N\ ,
\end{equation}
where $K$ is the maximal compact subgroup of $G(\Real)$ and we
have also indicated the Iwasawa decomposition of $G(\Real)$ into elements
of the maximal compact, Abelian and nilpotent subgroups $K,A,N$
respectively. This structure may describe only a part of the moduli space,
like the internal metric for toroidal compactifications or the Neveu-Schwarz
moduli for perturbative strings, or the full moduli space when
protected by supersymmetry.
Duality symmetries identify different points in this moduli space
and form an infinite discrete group $G(\Zint)$ of the classical
symmetry group $G(\Real)$, acting from the right on the coset $K\backslash
G(\Real)$.
The BPS states are labelled by a set of integer charges $m$ and fall into certain
representations $\R$ of the duality group. The mass of half-BPS states
is given by the manifestly $G(\Zint)$-invariant form
\begin{equation}
 \M^2 (\R) = m \cdot M_{\R}(g) \cdot m \ .
\end{equation}
Here, the moduli matrix $M_\R $ in representation $\R$ is given by
\begin{equation}
\label{moduli}
 M_{\R}(g) = \R^t \R (g)  \sp g \in A \cdot N \ ,
\end{equation}
where $g$ denotes an element in the coset $K \backslash G(\Real)$.
Moreover, there is in general a half-BPS constraint,
quadratic in the charges, that
can be written schematically as
\begin{equation}
\label{constraint} \mbox{1/2-BPS}: \qquad k = m \wedge m =0 \ ,
\end{equation} where the cup product $m\wedge m$ denotes all but
the highest of the irreducible components of the symmetric tensor
product $\R \otimes_s\R$. The constraint thus amounts to requiring
that $\R \otimes_s\R$ be an irreducible representation.

To be more specific, in the case of the $Sl(n,\Zint)$ duality
groups (of toroidally compactified diffeomorphism invariant
theories) we will be mostly interested in the fundamental and
antifundamental representations, labelled by the charges $m^i$ and
$m_i$ respectively. In this case the moduli matrix $M$ in
\eref{moduli} is simply the (inverse) metric $g_{ij}$ ($g^{ij})$
on the torus. For the T-duality group of toroidally compactified
string theory we will need the vector, spinor and conjugate spinor
representations, while for the U-duality group of M-theory the
particle, string and membrane representations will be of relevance
(see \cite{Obers:1998fb,Obers:1998rn,Obers:1999um} for details).

Given the duality group $G(\Zint)$, one may follow the  definition of Eisenstein
series given in the mathematical literature by \cite{Harish:1968}
\begin{equation}
\label{geneis2} {\cal E}^{G(\Zint)}_{\left\{w_i\right\} }(g) =
\sum_{h\in G(\Zint)/N} \prod_{i=1}^r  a_i( gh )^{-w_i}\ ,
\end{equation}
where $w_i$ is now an arbitrary $r$-dimensional vector ($r=\Rank
G$) in weight space, and $a(g)$ is the Abelian component of $g$ in
the Iwasawa decomposition \eref{iwasawa}. Note that this
definition is manifestly $K$-invariant on the left and
$G(\Zint)$-invariant on the right. This definition is however
hardly tractable, and we rather define our
generalized Eisenstein series \cite{Kiritsis:1997em,Obers:1999um} as
\begin{equation}
\label{geneis} \eis{G(\Zint)}{\R}{s}(g) = \sum_{m\in
\Lambda_\R\backslash\{0\} } \delta( m\wedge m) ~\left[ m \cdot
\R^t \R(g) \cdot m \right ]^{-s}
\end{equation}
for any symmetric space $K \backslash G(\Real)$, any representation $\R$ of
$G$ and any order $s$. This definition, albeit less general\footnote{Choosing $w$ along a
highest-weight vector $\lambda_{\R}$ associated to a
representation $\R$ reduces \eref{geneis2} to \eref{geneis} where
$w=s\lambda_{\R}$, up to an $s$-dependent factor.
These two definitions generalize the two terms of the last equality in
\eref{sl2eis} to higher rank groups.} than
\eref{geneis2}, has a more transparent physical meaning: the lattice
$\Lambda_\R$ labels the set of BPS states in the representation
$\R$ of the duality group, $\M^2=m \cdot \R^t \R(g) \cdot m $
gives their mass squared (or tension), and the $\delta$-function
imposes the half-BPS condition \eref{constraint}. This
ensures that the sum runs over one $G(\Zint)$ orbit only,
and is in fact a necessary requirement for the Eisenstein series to be an
eigenmode of the Laplacian on the symmetric space. The eigenvalue follows
simply from group theory,
\begin{equation}
\label{eisgen} \Delta_{K\backslash G}
\eis{G(\Zint)}{\irrep{\R_\lambda}}{s} = s (\lambda,\rho-s \lambda)
\eis{G(\Zint)}{\irrep{\R_\lambda}}{s} \ ,
\end{equation}
where $\lambda$ is the highest weight of the representation and
$\rho$ is the Weyl vector \cite{Obers:1999um}.
This expression implies in particular
that Eisenstein series associated to representations related by
outer automorphisms, i.e. symmetries of the Dynkin diagram, are
degenerate under $\Delta_{K\backslash G}$, as well as two
Eisenstein series of same representation but order $s$ and the
``dual'' value $[(\lambda,\rho)/(\lambda,\lambda)]-s$. Up to
these identifications, we expect that, aside from cusp forms, a
basis of invariant functions on the space is spanned by the
representations corresponding to the nodes of the Dynkin diagram.

\section{Perturbative string amplitudes}

\subsection{One-loop modular integral}
 Under toroidal compactification on a torus $T^d$, any
string theory exhibits the T-duality symmetry $SO(d,d,\Zint)$, and
all amplitudes should be expressible in terms of modular forms of
this group. For half-BPS saturated couplings, the one-loop
amplitude often reduces to an integral of a lattice partition
function over the fundamental domain $\F$ of the moduli space of
genus-1 Riemann surfaces,
\begin{equation}
\label{oneloop} I_d = 2\pi \int_\F \frac{d^2\tau}{\tau_2^2}
Z_{d,d} (g,B;\tau) \ ,
\end{equation}
where $Z_{d,d}$ is the partition function (or theta function) of
the even self-dual lattice describing the toroidal
compactification. This is for instance the case for $R^4$
couplings in type II strings on $T^d$, or $R^2$ or $F^2$ couplings
in type II on $K_3\times T^2$.

It is natural to expect a connection between this one-loop modular
integral and the $SO(d,d,\Zint)$ Eisenstein series defined above.
As is well known, the $\tau$-integral can be carried out by the
method of orbits, which corresponds to a large volume expansion of
the integral. This yields \cite{Kiritsis:1997em,Kiritsis:1997hf}
\begin{equation}
\label{largevolI}
 I_d = \frac{2\pi^2}{3}V_d +2V_d \eis{Sl(d,\Zint)}{\irrep{d}}{s=1}
 + \ldots
 \end{equation}
and exhibits the order 1 $Sl(d,\Zint)$ Eisenstein series in the
fundamental representation. The omitted terms are exponentially
suppressed world-sheet instanton terms that can be written down
exactly and simplify for the low-dimensional cases, reducing to the
standard results:
\begin{equation}
I_1 = \frac{2 \pi^2}{3} \left( R+ \frac{1}{R} \right)\ ,\qquad I_2
=-2\pi \log \left( T_2 U_2|\eta(T)\eta(U)|^4 \right)\ .
\end{equation}
Treating these simple cases first,  it is not difficult to show
that they can be rewritten as the sum of the $SO(d,d,\Zint)$
Eisenstein series of order 1 in the spinor and conjugate spinor
representations:
\begin{equation}
\label{cc1} I_d = 2 \eis{SO(d,d,\Zint)}{\spi}{s=1} + 2
\eis{SO(d,d,\Zint)}{\spb}{s=1} \sp d = 1,2\ .
\end{equation}
In particular, the result is manifestly invariant under the
extended T-duality $O(d,d,\Zint)$, where the extra generator
exchanges the two spinors.

To arrive at a similar claim for $d>2$, one considers the
eigenvalues  of the one-loop integral under both the Laplacian
$\Delta_{SO(d,d)}$ on the moduli space $SO(d)\times
SO(d)\backslash SO(d,d,\Real)$, as well as under another second order
differential operator $\square_d$. This operator is given by
\begin{equation}
\label{box} \square_d= \Delta_{Gl(d)} - \frac{1}{8} \left(%\sum_{ij}
g_{ij} \frac{\pa}{ \pa g_{ij} } \right)^2
=
\Delta_{Sl(d)} + \frac{2-d}{8d} \left( %\sum_{i j}
g_{ij} \frac{ \pa}{\pa g_{ij} } \right)^2
\end{equation}
which is non-invariant under $SO(d,d)$, but still invariant under
complete T-duality on all directions, and happens to have $I_d$ as
an eigenvector . The resulting eigenvalues are \cite{Obers:1999um}
\begin{equation}
\label{Ieigval}
 \Delta_{SO(d,d)}  I_d = \frac{d(2-d)}{4} I_d \sp
 \square_d   I_d = \frac{d(2-d)}{8} I_d \ .
\end{equation}
On the other hand,  the corresponding eigenvalues under these
operators
 of the
 Eisenstein series in the vector, spinor and conjugate spinor
representation are given by,
\begin{equation}
\label{soeigval} \Delta (\vect,s) = s (s-d +1) \ ,\qquad \Delta
(\spi,s) = \Delta (\spb,s) = \frac{s d (s-d +1)}{4}
\end{equation}
\begin{eqnarray}
\label{sqeigval} \square_d \eis{SO(d,d,\Zint)}{\vect}{s} &=&
 \frac{s(s-d+1)}{2} \eis{SO(d,d,\Zint)}{\vect}{s} \\
\square_d \eis{SO(d,d,\Zint)}{\spi}{s=1}& = & \frac{d(2-d)}{8}
\eis{SO(d,d,\Zint)}{\spi}{s=1}  \\ \square_d
\eis{SO(d,d,\Zint)}{\spb}{s=1}  &=&
\frac{d(2-d)}{8}\eis{SO(d,d,\Zint)}{\spb}{s=1} \ .
\label{bspieig}
\end{eqnarray}
%Note that for $s=1$ and $d=4$ the eigenvalues of the three
%representations are degenerate for both differential operators,
%and we conjecture that they are actually equated by triality.

Comparing \eref{Ieigval} with \eref{soeigval}-\eref{bspieig}, we
see that the candidate Eisenstein series for the one-loop integral
are restricted to the order $s=1$ spinor and conjugate spinor
series, together with the order $s=d/2-1$ vector series and their
duals. Another constraint comes from comparison of the large
volume expansion \eref{largevolI} with the expansion of the
Eisenstein series. These two requirements then enable one to show
that the one-loop integral $I_d$ in \eref{oneloop} can be
represented for $d \geq 3$ as
\begin{equation}
\label{c2} I_d = 2
\frac{\Gamma(\frac{d}{2}-1)}{\pi^{\frac{d}{2}-2}}
\eis{SO(d,d,\Zint)}{\vect}{s=\frac{d}{2}-1} = 2
\eis{SO(d,d,\Zint)}{\spi}{s=1} = 2 \eis{SO(d,d,\Zint)}{\spb}{s=1}
\ .
\end{equation}
Here, the first equality is a theorem and the last two are
well-supported conjectures \cite{Obers:1999um}.

As a practical application of this result, let us consider the
conjectured duality between the heterotic string on $T^4$ and type IIA
on $K_3$. On the heterotic side, the half-BPS states are momentum
and winding states, and transform as a vector of $SO(4,20,\Zint)$.
On the type IIA side at the $T^4/\Zint_2$ orbifold point, these
states correspond instead to even branes wrapped on invariant
cycles of $T^4$, plus ``fractional'' branes stuck at the fixed
points of the orbifolds. Restricting to an $[SO(4)\times
SO(4)]\backslash SO(4,4)$ subspace of the moduli space, it is easy
to see that under duality, the vector of $SO(4,4)$ should be
mapped to the conjugate spinor: the duality therefore corresponds
to a triality in $SO(4,4)$, and using our last result \eref{c2},
it is easy to see that $I_4$ is indeed invariant under
heterotic-type IIA duality. This fact will be instrumental in
checking the duality conjecture at the level of $F^4$ couplings
\cite{Kiritsis:1999}.

\subsection{Genus $g$ modular integral}

Similar methods carry over to higher-loop amplitudes, such as the
higher-genus analogue of \eref{oneloop}, namely the integral of a
lattice partition function on the $3g-3$-dimensional moduli space
${\cal{M}}_g$ of genus $g$ curves
\begin{equation}
\label{gloop} I^g_d=\int_{{\cal{M}}_g}  d\mu~ Z_{d,d}^g
\left(g_{ij},B_{ij};\tau\right) \ ,
\end{equation}
where $Z_{d,d}^g$ is the genus $g$ lattice sum.  In this case the
modular group is  $Sp(g,\Zint)$ and one may derive by similar
methods as in the one-loop case the eigenvalue condition,
\begin{equation}
\label{eighg} \Delta_{SO(d,d)} I_d^g = \frac{d g(g+1-d)}{4} I_d^g
\ .
\end{equation}
Comparison with \eref{soeigval} shows that this eigenvalue agrees
with the order $s=g$ Eisenstein series in the spinor and conjugate
spinor representation. This leads us to the conjecture that the
$g$-loop integral \eref{gloop} is (up to an overall factor) given
by the $SO(d,d,\Zint)$ Eisenstein series of order $g$ in the
spinor representation:
\begin{equation} \label{c3} I_d^g \propto
\eis{SO(d,d,\Zint)}{\spi}{s=g} + \eis{SO(d,d,\Zint)}{\spb}{s=g} \
,
\end{equation}
where the superposition of the two spinor representations is
required by the $O(d,d,\Zint)$ invariance of the integrand. Note
that normalizing \eref{c3} would require a knowledge of the
Weil-Peterson volume of the moduli space of genus $g$ curves.

Though less substantiated than  the one-loop conjecture \eref{c2},
\Eref{c3} is strongly reminiscent of the genus $g$ partition
function of the $N=4$ topological string \cite{Berkovits:1995vy}
on $T^2$ which was shown to be exactly given by the Eisenstein
series of order $s=g$ in the spinor representation
$\eis{Sl(2,\Zint)}{2}{s=g}(T)$~\cite{Ooguri:1995cp}. This result
was subsequently used to derive a set of higher derivative
topological couplings $R^4 H^{4g-4}$ in type IIB string
compactified over $T^2$ \cite{Berkovits:1998ex}. Our conjecture
\eref{c3} suggests a natural generalization of these results to
lower dimensions: The $R^4 H^{4g-4}$ couplings between 4 gravitons
and $4g-4$ Ramond three-form  field-strengths in type IIA
compactified on $T^d$, $d\leq 4$ are given at genus $g$ by the
$SO(d,d,\Zint)$ constrained Eisenstein series in the spinor
representation with insertions of $4g-4$ charges:
\begin{equation}
\label{rhp} I=\int d^{10-d} x \sqrt{-\gamma}\ \hat{\sum_{m}}
\delta(m\wedge m)\ e^{6(g-1)\phi} \ \frac{R^4 \ (m\cdot
H_{RR})^{4g-4}} {\left(m\cdot {M}(\irrep{\spi}) \cdot
m\right)^{3g-2}} \ ,
\end{equation}
where $\phi$ is the T-duality invariant dilaton, related to the
ten-dimensional coupling as $e^{-2\phi}=V_d/g_s^2 l_s^d$, and we
work in units of $l_s$. \
%The restriction $d\leq 4$ is due to the
%fact that for $D = 5$ three-form field-strength are Poincar{\'e}
%dual to two-form field-strengths, while for $D=4$ they become part
%of the scalar manifold after dualization.
A similar conjecture
also holds for the coupling computed by the topological B-model
\cite{Berkovits:1995vy}, and states that the $R^4 F^{4g-4}$
couplings between 4 gravitons and $4g-4$ Ramond two-form
field-strengths in type IIA compactified on $T^d$, $d\leq 6$  are
given at genus $g$ by the $SO(d,d,\Zint)$ constrained Eisenstein
series in the conjugate spinor representation with insertions of
$4g-4$ charges:
\begin{equation}
\label{rfp} I=\int d^{10-d} x \sqrt{-\gamma} \ \hat{\sum_{m}}
\delta(m\wedge m) \ e^{6(g-1)\phi} \frac{R^4 \ (m\cdot
F_{RR})^{4g-4}} {\left(m\cdot {M}(\irrep{\spb}) \cdot
m\right)^{3g-2}} \ .
\end{equation}
%Here, the restriction $d\leq 6$ is due to the fact that for $D =
%3$ two-forms field-strengths become part of the scalar manifold
%after Poincar{\'e} dualization.
Note that the results \eref{rhp} and \eref{rfp} involve
covariant modular functions instead of invariant ones, but behave
like Eisenstein series of order $3g-2-(4g-4)/2=g$ as far as their
eigenvalue and decompactification limit are concerned.
They generalize the $Sl(2,\Zint)$ modular functions
$f^{p,q}={\hat{\sum}}
\tau_2^{(p+q)/2}/[(m+n\tau)^p(m+n\bar\tau)^q]$ invariant up to a
phase, that were also used in the context of non-perturbative type
IIB string in \cite{Kehagias:1997cq,Green:1998me}.

\section{Non-perturbative string amplitudes}
\subsection{$R^4$ couplings in toroidally compactified type II}

While Eisenstein series provide a nice way to rewrite one-loop
integrals such as \eref{oneloop}, their utility becomes even more
apparent when trying to extend the perturbative computation into a
non-perturbatively exact result. Indeed, a prospective exact
threshold should reduce in a weak coupling expansion to a sum of
T-duality invariant Eisenstein-like perturbative terms, plus
exponentially suppressed contributions, and Eisenstein series of
the larger non-perturbative duality symmetry are natural
candidates in that respect.

Four graviton $R^4$ couplings in maximally supersymmetric theories
have been argued in dimension $D\ge 8$ to receive no perturbative
corrections beyond the tree-level and one-loop terms, and to be
eigenmodes of the Laplacian on the full scalar manifold $K
\backslash \exc(\Real)$ as a consequence of supersymmetry. We
assume both properties to persist in lower dimensions as well, so
that from the first property we have
\begin{equation}
\label{pertr4} f_{R^4}= 2\zeta(3) \frac{V_d}{g_s^2} + I_d +
\mbox{non pert.}
\end{equation}
This result can be suggestively rewritten in terms of T-duality
Eisenstein series as,
\begin{equation}
\label{pertr4e}
 f_{R^4}= \frac{V_d}{g_s^2}
\eis{SO(d,d,\Zint)}{\irrep{1}}{s=3/2} +
\eis{SO(d,d,\Zint)}{\irrep{\spi}}{s=1} + \mbox{non pert.}
\end{equation}
While the Ramond scalars are decoupled from the perturbative
expansion by Peccei-Quinn symmetries, the full non-perturbative
result should depend on all the scalars in the symmetric space
$K\backslash \exc(\Real)$. The $R^4$ threshold should thus be an
automorphic form of $\exc(\Zint)$ with asymptotic behavior as in
\eref{pertr4e}. To obtain the non-perturbative extension of this
result, we use the fact that the singlet and spinor
representations of the T-duality group are unified into the string
representation \cite{Elitzur:1997zn,Obers:1997kk,Obers:1998fb} of
the U-duality group, the tension of which is given by
\begin{equation}
\T^2= m^2 + \frac{V_d}{g_s^2} \M^2(\spi) + \frac{V_d^2}{g_s^4} \ ,
\M^2(\irrep{O}) \ ,
\end{equation}
where $\irrep{O}$ is an extra representation that appears when $d
\geq 4$. From this group theory fact, one arrives at the
non-perturbative generalization of \eref{pertr4e}: The exact
four-graviton $R^4$ coupling in toroidal compactifications of type
II theory on $T^d$, or equivalently M-theory on $T^{d+1}$, is
given, up to a factor of Newton's constant, by the Eisenstein
series of the U-duality group $\exc(\Zint)$ in the {\it string
multiplet} representation, with order $s= 3/2$:
\begin{equation}
\label{cr4} f_{R^4} = \frac{V_{d+1}}{l_M^9}
\eis{\exc(\Zint)}{\irrep{string}}{s=3/2} \ .
\end{equation}
Here $l_M$ is the eleven-dimensional Planck length, $V_{d+1}=R_s
V_d$ the volume of the M-theory torus $T^{d+1}$. The quantity
$V_{d+1}/l_M^9=l_P^{d-8}$ is the U-duality invariant gravitational
constant in dimension $D=10-d$. As an immediate check, the
proposal has the appropriate scaling dimension $d+1-9+3\times 2$
for an $R^4$ coupling in dimension $D=10-d$.

Using the eigenvalue equation \eref{eisgen} we can in fact
determine the corresponding eigenvalues of \eref{cr4} under the
Laplacian on the symmetric space $K \backslash\exc(\Real)$
\begin{equation}
\label{npeig} \Delta_{\exc} f_{R^4} = \frac{3(d+1)(2-d)}{2(8-d)}
f_{R^4}\ .
\end{equation}
This property could in principle be proved from supersymmetry
arguments along the lines of \cite{Pioline:1998mn,Green:1998by},
and holds order by order in the the weak coupling expansion. In
fact, using the eigenvalues of the particle and membrane
representations of the U-duality group one arrives at the
conjecture that the non-perturbative $R^4$ amplitude in M-theory
on $T^{d+1}$ has three equivalent forms,
\begin{equation}
\label{ccc} \fl 
\frac{V_{d+1}}{l_M^9}\eis{\exc(\Zint)}{\irrep{string}}{s=3/2}
=\frac{\Gamma(d/2-1)}{\pi
^{d/2-2}}\eis{\exc(\Zint)}{particle}{s=d/2-1}=
\frac{V_{d+1}}{l_M^9}\eis{\exc(\Zint)}{\irrep{membrane}}{s=1} \ .
\end{equation}
Again, it is easy to check that the scaling dimensions match.

As a justification of the claim \eref{cr4}, it can be shown that
it reproduces the perturbative contributions in \eref{pertr4} in a
weak coupling expansion. Moreover, taking the $d=4$ case as an
example  the non-perturbative terms can be interpreted as
superposition of Euclidean D0 and D2-branes wrapped on a one-cycle
 or a three-cycle of $T^4$. In addition to these terms, there are further
contributions which behave superficially as $e^{-1/g_s^2}$. Such
non-perturbative effects are certainly unexpected in toroidal
compactifications to $D>4$, since there are no half-BPS instanton
configurations with this action (the NS5-brane does have a tension
scaling as $1/g_s^2$, but it can only give rise to Euclidean
configurations with finite actions when $D\leq 4$). The matching
of the tree-level and one-loop contributions together with the
consistent interpretation of the D-brane contribution is however a
strong support to our conjecture.

\subsection{$R^4 H^{4g-4}$ and $R^4 F^{4g-4}$ couplings}

Using the manifestly T-duality invariant forms \eref{rhp} and
\eref{rfp} of the $g$-loop amplitudes, it is also straightforward
to propose a non-perturbative completion of the $R^4 H^{4g-4}$ and
$R^4 F^{4g-4}$ couplings, invariant under the full U-duality
group. For that purpose, we note that the set of three-form
field-strengths in M-theory compactified on $T^{d+1}$ fall into a
representation of {\exc} dual to the string multiplet.
%(this is strictly speaking only
%correct for $D \geq 5$ as explained \eref{rhp}).
The string multiplet decomposes under $SO(d,d,\Zint)$ into a
singlet (the Neveu-Schwarz $H_{NS}$), a spinor (the Ramond
three-forms obtained by reducing the M-theory four-form
field-strength), as well as further terms for $d\ge 4$. We
therefore conjecture that the $R^4 H^{4g-4}$ couplings between 4
gravitons and $4g-4$ three-form field-strengths in M-theory
compactified on $T^{d+1}$, $d\leq 4$ are exactly given, up to a
power of Newton's constant, by the $\exc(\Zint)$ constrained
Eisenstein series in the string representation with insertions of
$4g-4$ charges:
\begin{equation}
\label{rh} I=\frac{V_{d+1}}{l_M^9} \int d^{10-d} x \sqrt{-\gamma}\
\hat{\sum_{m}} \delta(m\wedge m)\ \frac{R^4\ (m\cdot H)^{4g-4}}
{\left(m\cdot {M}(\irrep{string}) \cdot m\right)^{3g-\frac{3}{2}}}
\ .
\end{equation}
As an immediate check, we note that this proposal has the
appropriate scaling dimension, while the expression also
reproduces the tree-level interaction involving the Neveu-Schwarz
three-form only. Moreover, the result \eref{rh} reproduces the
$g$-loop result \eref{rhp}. The analysis of non-perturbative
effects is as in the $R^4$ case, and shows order $e^{-1/g_s}$
D-brane effects as well as, for $d\ge 4$, contributions
superficially of order $e^{-1/g_s^2}$.

Similarly, for the case of non-perturbative $R^4 F^{4g-4}$
couplings, we note that the two-form field-strengths of M-theory
compactified on $T^{d+1}$ transform as the dual of the particle
multiplet. This makes it natural to propose that
\begin{equation}
\label{rf} I=\int d^{10-d} x \sqrt{-\gamma}\ \hat{\sum_{m}}
\delta(m\wedge m)\ \frac{R^4\ (m\cdot F)^{4g-4}} {\left(m\cdot
{M}(\irrep{particle}) \cdot m\right)^{4g-5+\frac{d}{2}} } \ ,
\end{equation}
where the power $4g-5+d/2$ has been set by dimensional analysis.
The particle multiplet decomposes as a vector and conjugate spinor
of $SO(d,d,\Zint)$ in that order, so that this proposal implies a
one-loop term given by the $SO(d,d,\Zint)$ Eisenstein series of
order $2g-3+d/2$ in the vector representation, plus a higher
perturbative term which should reproduce the genus $g$ term
\eref{rfp}. Due to the presence of constraints, this statement has
not been proved at present.

\section{Open directions}

The analysis presented here has focussed on half-BPS saturated
couplings in theories with maximal supersymmetry. It would be
interesting to extend our techniques to (i) couplings preserving a
lesser amount of supersymmetry, and (ii) half BPS states in
theories with less supersymmetry. Given that the quadratic
half-BPS constraint imposes second order differential equations
and that the quarter-BPS condition is cubic in the charges, one
may envisage that quarter-BPS saturated couplings should be
eigenmodes of a cubic Casimir operator, and expressable as
generalized Eisenstein series with quarter-BPS conditions
inserted.

On a more mathematical level, our results provide a wealth of
explicit examples of modular functions on symmetric spaces of
non-compact type $K\backslash G$, with $G$ a real simply laced Lie
group in the normal real form, that generalize the Eisenstein
series on the fundamental domain of the upper half-plane. We have
not addressed the question of the analyticity of Eisenstein series
with respect to the order $s$. Unfortunately, the
presence of constraints tends to give rise to ill-behaved
expansions, which is  the mathematical counterpart of the physical
problem raised above of understanding the instanton effects
superficially of order $e ^{-1/g_s^2}$.

It would also be interesting to understand more precisely what
Eisenstein series are needed to generate the spectrum of the
Laplace operator for any eigenvalue (note in that respect that the
order $s$ is no longer a good parametrization, since the relation
between the eigenvalue and $s$ depends on the representation).
From the point of view of harmonic analysis however, Eisenstein series are
the least interesting part of the spectrum on such manifolds,
which should also include a discrete series of cusp forms. Hopefully
string theory will provide an explicit example of these elusive
objects.

\section*{References}

%\bibliographystyle{h-elsevier}
%\bibliography{biblioniels}
%\end{document}

\end{document}